# SOCIAL CORRECTION ON SOCIAL MEDIA: A QUANTITATIVE ANALYSIS OF COMMENT BEHAVIOUR AND RELIABILITY

*Completed Research Paper*

Sameera S. Vithanage, School of Computing and Information Systems, University of Melbourne, Melbourne, Australia, sameera.vithanage@student.unimelb.edu.au

Keith Ransom, School of Computer and Mathematical Sciences, University of Adelaide, Adelaide, Australia, keith.ransom@adelaide.edu.au

Antonette Mendoza, School of Computing and Information Systems, University of Melbourne, Melbourne, Australia, mendozaa@unimelb.edu.au

Shanika Karunasekera, School of Computing and Information Systems, University of Melbourne, Melbourne, Australia, karus@unimelb.edu.au

## Abstract

*Corrections given by ordinary social media users, also referred to as Social Correction have emerged as a viable intervention against misinformation as per the recent literature. However, little is known about how often users give disputing or endorsing comments and how reliable those comments are. An online experiment was conducted to investigate how users' credibility evaluations of social media posts and their confidence in those evaluations combined with online reputational concerns affect their commenting behaviour. The study found that participants exhibited a more conservative approach when giving disputing comments compared to endorsing ones. Nevertheless, participants were more discerning in their disputing comments than endorsing ones. These findings contribute to a better understanding of social correction on social media and highlight the factors influencing comment behaviour and reliability.*

*Keywords: Social media, Misinformation, Social Correction*

# 1      Introduction

The emergence of social media has significantly altered the communication landscape by making it easier to publish and disseminate information quickly and widely and at a low cost. While timely and rapid information sharing may enhance social relationships, it also provides ample opportunity to disseminate and amplify misinformation. The existing literature classifies false content on social media using various terms (including fake news, disinformation, and misinformation) based on the intent or the editorial process followed when creating and sharing information. However, the current study focuses on such content not from the creator's point of view but from the receiver's. Hence, we adopt the broader definition of misinformation as "*cases in which people's beliefs about factual matters are not supported by clear evidence and expert opinion*" (Nyhan and Reifler, 2010, p.305). Misinformation has been found to have far-reaching and detrimental consequences in different aspects of society. In the political sphere, it can significantly influence the outcomes of democratic processes, potentially undermining the legitimacy of governing institutions (Allcott & Gentzkow, 2017; Bovet & Makse, 2019). Economically, misinformation has been shown to damage brand reputations (Mikkelson, 2015) and disrupt financial markets (ElBoghdady, 2023). Moreover, the public health sector is also vulnerable to the detrimental effects of misinformation. For instance, it has been identified as a major contributor to vaccine hesitancy (Basch et al., 2021; Calo et al., 2021; Wu et al., 2022) and has even led to tragic outcomes, such as the hundreds of deaths caused by the false belief that ingesting methanol could prevent COVID-19 (Heidari & Sayfouri, 2021).

The severity of such consequences has drawn a considerable amount of scholarly attention aimed at minimising the spread of misinformation and mitigating its harmful effects. Numerous interventions have been proposed and tested in current literature, ranging from algorithmic solutions such as machine learning models for misinformation detection to psychological interventions such as inoculation (Linden & Roozenbeek, 2020). One widely proposed intervention is correction, which recent meta-analyses suggest is effective against misinformation (Walter et al., 2021; Walter & Murphy, 2018). Corrections to misinformation posts can originate from two primary sources. Professional entities, such as journalists, fact-checking organisations and government organisations may actively post corrections to misinformation as part of their roles. Similarly, ordinary social media users may also provide corrections to misinformation posts they encounter – a process known as *social correction* (Kligler-Vilenchik, 2022; Sun, Chia, et al., 2022; Sun, Oktavianus, et al., 2022). Although corrections from professionals are known to reduce misperceptions (believing in misinformation), a limitation inherent in this approach is that no fact-checking team or editorial team has the capacity to address the sheer volume of misinformation on social media directly (Koo et al., 2021; Micallef et al., 2020). In contrast, social correction may not face this capacity constraint. A study of corrections to COVID-19 misinformation on Twitter found that 96% originated from ordinary users, and these social corrections received more online attention than those given by professional fact-checkers (Micallef et al., 2020).

Social media users typically engage in social correction through two distinct approaches: posting comments that dispute the credibility of a post when they perceive it as containing misinformation or creating independent posts on their own social media accounts to address false information that is being disseminated online (Bautista et al., 2021a, 2023). In this study, we focus on comments given by social media users. Prior to commenting, users would assess if the post contains true information or misinformation. This credibility evaluation will likely influence the decision to provide a comment as well as what type of comment is given. Based on the user's credibility evaluation, we identify two types of comments: disputing and endorsing. Specifically, disputing comments challenges the credibility of a post perceived as false, although users may also incorrectly dispute true posts. Consequently, for a disputing comment to be considered a correction, the user's initial credibility evaluation must accurately identify the post as containing misinformation. Similarly, if the user believes a post is truthful, they might give endorsing comments affirming their belief in the claim.

The effectiveness of social feedback - both disputing and endorsing hinges on several factors, including the frequency and reliability of the comments. Research has consistently demonstrated that social media users often rely on comments posted by others when evaluating the credibility of content (Bryanov &

Vziatysheva, 2021; Colliander, 2019; Geeng et al., 2020; Ransom et al., 2021; Wijenayake et al., 2021). It has been found that witnessing disputing comments given by other users can be as effective in reducing misperceptions as corrections from reputed government organizations such as the U.S. Centres for Disease Control and Prevention (CDC) or those issued directly by social media platforms (Bode & Vraga, 2018; Sullivan, 2019; Vraga & Bode, 2017). In contrast, however, disputing comments on posts that carry true information can lead people to distrust facts, while endorsing comments on misinformation can lead them to believe it (Vraga & Bode, 2022; Wijenayake et al., 2021). Therefore, it is crucial that users provide comments that endorse information when it is true or factual and dispute it when it is not. Ideally, the highest proportion of disputing and endorsing comments should come from users with accurate credibility evaluations, yet sometimes, the comments may come from users who are entrenched in misperceptions. For example, when giving comments related to the consumption of raw milk, individuals who held misperceptions were more likely to post comments online compared to those who held correct beliefs (Tully et al., 2020). Similarly, those who held misperceptions about COVID-19 were more likely to dispute others (Bode & Vraga, 2021).

Maintaining a positive impression online is one of the key motivators behind social media use (Hollenbaugh, 2021). Consequently, users are likely to consider the potential impact of their comments on their own reputation, as well as that of the original poster, given the broad audience reach of online content. Disputing a post publicly may have reputational implications for the original poster, and if the user mistakenly disputes a true post, it may even backfire on them. Conversely, there can be reputational costs associated with endorsing misinformation. Thus, reputational concerns may influence their commenting behaviour in conjunction with credibility evaluations.

Even though prior research indicates that disputing comments on misinformative posts can mitigate misperceptions, significant gaps remain in our understanding of user commenting behavior. Specifically, studies have not examined how often users give such comments compared to endorsing comments. Furthermore, the accuracy of users' initial credibility assessments, which inform their comments, is unknown. This raises questions about the extent to which individuals can rely on user comments for accurate credibility evaluation. Therefore, to address this gap, we conducted an exploratory study to investigate how users' confidence in their credibility evaluations and their online reputation concerns influence the relative frequency and reliability of the comments they give. The study found that despite users being biased towards perceiving social media posts to be false more often than true, they relatively gave fewer disputing comments compared to endorsing ones. Nevertheless, the disputing comments they gave were generally more reliable than endorsing ones.

## 2  Literature Review

### 2.1  Impression Management on Social Media

Social media platforms were primarily designed to facilitate individuals to interact with each other by creating and sharing content. However, online interactions mirror offline ones in that individuals consistently consider how their actions are perceived by others and the impressions they form (Goffman, 1959). This phenomenon is formally known as impression management or self-presentation, defined as *"the process by which individuals attempt to control the impressions others form of them"* (Leary & Kowalski, 1990, p. 34). According to Leary and Kowalski's Two-Component Model of Impression Management (1990), individuals engage in impression management to achieve three interconnected objectives: acquiring social and material rewards, enhancing self-esteem, and developing and maintaining desired identities.

The Two-Component Model further posits that impression management comprises two interconnected processes: impression motivation and impression construction. Impression motivation refers to the degree to which an individual is driven to manage their impression, while impression construction encompasses the behavioural actions taken to achieve a desired impression (Leary & Kowalski, 1990). A primary factor that affects impression motivation is the degree of publicity surrounding an individual's action, with larger potential audiences increasing motivation to manage impressions. In the context of social media, the content a user posts has the potential to reach a wide audience, which can include close social ties (e.g. Family and friends) as well as distant ones (e.g. work colleagues).

Moreover, a factor that is not directly relevant to offline settings and, therefore, not considered in earlier self-presentation theories but applies in online settings is persistence. Unlike offline interactions, social media content and subsequent feedback given by other users persist online unless explicitly removed (Hollenbaugh, 2021). Consequently, individuals may be more motivated to manage their impressions on social media than in offline settings, particularly in day-to-day interactions, due to the enduring nature of online content as well as the potential to reach a wider audience. Consistent with this notion, research indicates that social media users strategically present themselves to garner positive impressions. Studies have shown that users selectively post content they believe will be perceived favourably by their audience and sometimes employ measures such as manipulating the images they post to achieve this goal (B. K. Johnson & Ranzini, 2018, 2018; Lyu, 2016; Roulin & Levashina, 2016). Extending beyond individual users, organisational entities also engage in impression management on social media to foster positive perceptions among their target audience (Benthaus et al., 2016; Yang & Liu, 2017).

Given that impression construction refers to actions an individual would take to make an impression on others, in the context of our study, this would amount to providing disputing or endorsing comments. We believe that the potential to achieve the goals of portraying desired identities, gaining rewards, and gaining self-esteem would motivate users to manage their impressions when giving comments on social media and this, in turn, would affect their commenting behaviour.

## 2.2 Hypothesis Development

### 2.2.1 Commenting on Social Media

The motivation to manage one's impressions is largely contingent upon the perceived value of the outcomes associated with such management (Leary & Kowalski, 1990). In the context of social media comments, these outcomes can be contextualised within a framework that considers both the user's credibility evaluation and the veracity of the post they are responding to, as illustrated in Figure 1. As mentioned earlier, users might give endorsing comments (quadrants A and B) when they perceive a post contains true information. However, it is unlikely that they would explicitly refer to the credibility of the post in such a comment. Instead, such comments would often express their opinion related to the content of the post, often aligning with their desired online identities. For instance, commenting on a political post might serve to assert one's political stance, while commenting on a climate change post could signal environmental consciousness. Further, they may expect to gain social capital by engaging with other users and even expect positive feedback from others, which will enhance their self-esteem (Burrow & Rainone, 2017).

|  | Ground truth of the post | |
|---|---|---|
|  | *True* | *False* |
| **Comment type** *Endorse* | A | B |
| **Comment type** *Dispute* | C | D |

*Figure 1.   The quadrant of outcomes is based on the ground truth of the original post and the type of comment given by the user.*

Unlike endorsing comments, disputing comments (quadrants C and D) would most likely directly challenge the accuracy of the post. By doing so, social media users might try to demonstrate that they are knowledgeable in the topic area (Kligler-Vilenchik, 2022). Moreover, they may be motivated by altruistic concerns, such as protecting others from misinformation (Gurgun et al., 2024), which would enhance their self-esteem. Despite the intention to attain a favorable impression from other users, disputing comments, at times, can have unintended negative consequences. By challenging the accuracy of another user's content, the commenter risks being perceived as a troublemaker or a critic (T. Johnson & Kromka, 2023), potentially eliciting critical responses to the disputing comment they gave (Gurgun et al., 2024) , which may reduce self-esteem. This potential for negative outcomes may influence users' perception of the value of disputing comments as a means of impression management. The higher likelihood of negative interpretations associated with disputing comments may diminish their perceived utility compared to endorsing comments. Consequently, we predict that users will exhibit lower motivation to engage in disputing comments relative to endorsing ones, given the increased risk of adverse social consequences.

Therefore, we hypothesize:

> *H1: Social media users will post fewer disputing comments for claims they perceive to be false than endorsing comments for claims they perceive to be true.*

Individuals, when engaging in impression construction, would consider the likelihood of achieving the desired social impression via the chosen behaviour (Leary & Kowalski, 1990). Therefore, before providing either a disputing or an endorsing comment, users will consider if they can achieve the intended impression they want to achieve by commenting. While social media users may intend to present positive impressions of themselves to others by providing comments, doing so may yield negative outcomes if they accidentally give a comment that mismatches the ground truth of the post they are commenting on. Social media users avoid sharing posts that they think contain misinformation as it would harm their online reputation (Altay et al., 2020). In the same vein, giving an endorsing comment on a post that carries misinformation (quadrant B) may lead the user to be perceived as gullible or naïve among their network, as they may be seen as lacking discernment or correct judgement.

Disputing comments, on the other hand, directly challenge the credibility of a post by highlighting inaccuracies. Even though giving a disputing comment to a truthful post (quadrant C) can risk much the same negative perception as the user being perceived as someone who lacks judgment, it may also carry additional risks. Previous studies have found that social media users tend to think that correcting others is somewhat aggressive behaviour as it looks like accusing another user publicly of spreading misinformation (Colliander, 2019; Tandoc et al., 2020). However, if the correction is given to a truthful post, the corrector himself may be perceived as the person spreading falsehoods, which might lead to critical responses from other users for attempting to harm another user's online reputation. Therefore, whether the user thinks the post they are commenting on is true or false and their confidence in this evaluation will likely play a role in the impression construction process as it may be directly related to the possibility of achieving the intended impression from the audience. We predict that users will comment only when they are sufficiently confident in their credibility evaluation – especially when correcting.

Therefore, we hypothesize:

> *H2: Higher confidence thresholds will be observed in social media users' credibility evaluations when:*
>> *a. they choose to comment on a claim compared to when they choose not to.*
>> *b. they give disputing comments compared to when they endorse the claim.*

### 2.2.2  Reliability of Comments on Social Media

Considering that there can be negative consequences for endorsing misinformation (quadrant B) and disputing true information (quadrant C), users would naturally want to minimise instances where the comment doesn't correspond to the ground truth (*misses*). Similarly, they would want to improve their chances of comments matching the ground truth (*hits*) such that they endorse true information (quadrant A) and dispute misinformation (quadrant D). Drawing from signal detection theory (SDT), this can be achieved through two pathways: improving the truth discernment ability of users to increase the accuracy of credibility evaluations, which requires cognitive effort, or adjusting their response bias (criterion) to optimise hits and minimise misses (Green & Swets, 1966).

Since humans tend to be cognitive misers that resist expending cognitive effort (Simon, 1979), especially in the case of social media (Pennycook & Rand, 2019), it is likely that users would rely on criterion shifts to maximise their chances of hits and minimise misses instead of spending cognitive effort on their commenting decisions.

Typically, individuals adopt a conservative criterion when the consequences of a miss are severe in order to minimise misses and a more liberal approach when the cost of a miss is relatively low (Lynn & Barrett, 2014). Given the potentially higher reputational cost associated with a missed disputing comment compared to a missed endorsing comment, we hypothesised earlier that users would give a relatively smaller number of disputing comments and demonstrate higher confidence thresholds when providing such comments. This suggests that users would employ a more conservative approach when offering disputing comments to minimise misses, in contrast to their behaviour when providing endorsing comments. Therefore, we predict that there will be relatively more hits compared to misses in the disputing comments than endorsing comments.

Therefore, we hypothesize,
> *H3: Social media users' disputing comments will more often correspond to claims that are actually false than endorsing comments that correspond to claims that are actually true.*

## 3    Method

The current study focused on investigating how often social media users give disputing comments compared to endorsing comments and how reliable those comments are. To that end, an online experiment was conducted, which was pre-registered at *https://aspredicted.org/Y2M_6HW*.  The study was approved by the University of Melbourne Human Ethics Advisory Group (ID: 25436).

### 3.1    Stimuli

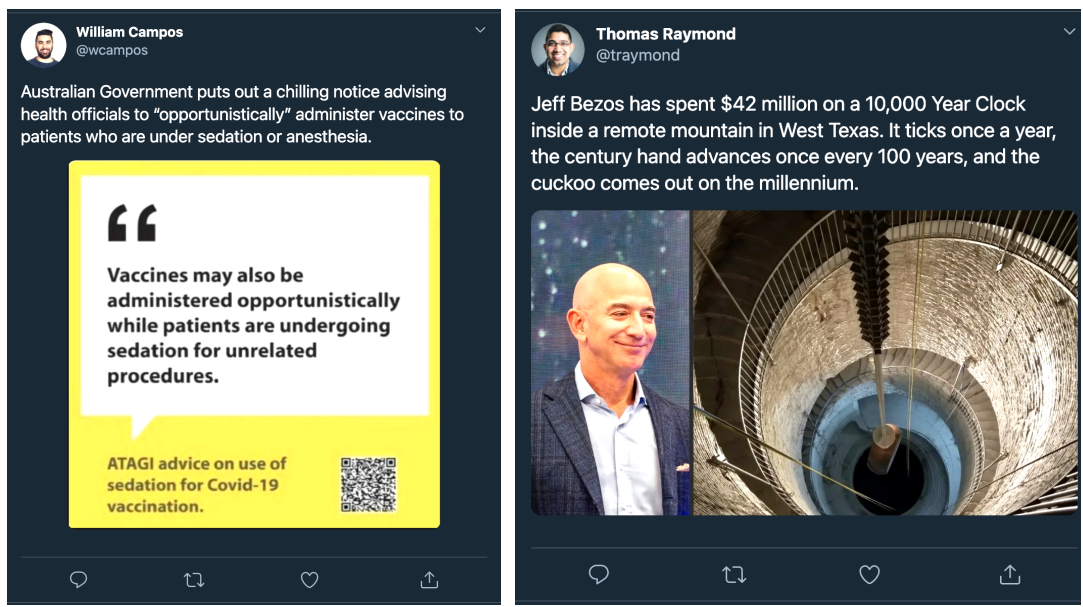

*Figure 2.        Sample stimuli – the first image contains an example of a misinformation post. The second image is an example of a post containing a true claim.*

Social media users tend to engage with content that is of personal interest to them (Tandoc et al., 2020; Wintterlin et al., 2021; Yaqub et al., 2020). Hence, a preliminary study was conducted to select a set of suitable stimuli to mitigate the risk of a floor effect where experiment participants refrain from commenting on any of the presented stimuli. An initial pool of 40 claims, evenly divided between true and false claims, was extracted from reputable fact-checking sites such as Snopes, PolitiFact, AFP fact checks and several mainstream media outlets in Australia. Eight undergraduate students from an Australian university participated in this prelim study, evaluating each claim in the format of a tweet (now referred to as X posts). Participants were asked to indicate whether they would comment on the claim if encountered online. The top ten claims from each category—misinformation and true posts—with the highest likelihood of eliciting comments were selected for the main experiment. Figure 1 depicts two examples from the selected set of stimuli.

## 3.2 Participants

A total of 150 undergraduate students from an Australian university participated in the main experiment. The average age of the sample was 19.29 years. The majority of participants (68.67%) identified as female, while approximately 25% identified as male. Notably, over 90% of participants reported daily use of social media platforms. A summary of these demographic details is presented in Table 1.

| | |
|---|---|
| **Age** | |
| Mean (SD) | 19.29 (1.91) |
| Minimum - Maximum | 18 - 30 |
| **Gender** | |
| Female | 103 (68.67%) |
| Male | 39 (26%) |
| Other | 8 (5.33%) |
| **Social Media Use** | |
| Daily | 139 (92.66%) |
| Weekly | 7 (4.66%) |
| Monthly | 1 (0.66%) |
| Rarely | 1 (0.66%) |
| Never | 1 (0.66%) |
| 'Prefer not to say' | 1 (0.66%) |

*Table 1.     Summary of the demographic data of the 150 participants in the main experiment.*

## 3.3 Design

After providing informed consent and passing a simple attention/comprehension check, participants completed a demographic questionnaire before proceeding to the test phase comprising 20 trials, shown in randomized order. On each trial, people were presented with a single tweet and asked to mention whether they thought the claim in the tweet was likely to be true or false, and to rate their confidence in their judgement on a scale of 0-100. Those who indicated that the claim was likely to be true were asked to provide an endorsing comment, while those who said it contained misinformation were asked to give a disputing comment. Lastly, participants were asked if they would actually post the comment online had they encountered this tweet. The study took thirty minutes to complete on average, and participants were awarded course credits for participation.

The resulting dataset comprised 3000 trial records (150 participants x 20 claims). To mitigate the impact of inattentive responses, trials with comments shorter than five characters or duration less than one-third of the median duration of 52 seconds were excluded following Freiling & Matthes's (2023) approach. This criterion aimed to remove instances of speeding where participants may have responded without reading the claim or gave responses without considering the claim, thereby reducing noise and improving data quality (Greszki et al., 2015). Consequently,158 records (5.26% of the entire dataset) were removed, resulting in a final dataset comprising 2842 records.

## 4 Results

The first hypothesis predicted that the proportion of perceived false claims that receive disputing comments by users would be less than the proportion of perceived true claims that receive endorsing comments. Figure 3 illustrates the distribution of the proportion of perceived true claims that were endorsed and perceived false claims that were disputed online. A two-sample test for equality of proportions with continuity correction revealed a significant difference between the two proportions *($\chi^2$ (1) = 7.01, p = .008)* such that a higher proportion of perceived true posts received endorsing comments online (25.67%) compared to the proportion of perceived false claims that received corrections (21.39%). Thus, supporting the first hypothesis.

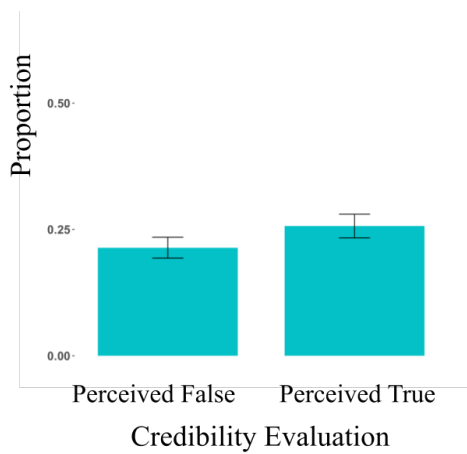

*Figure 3.    Distribution of the proportion of the claims that received comments online between the two types of credibility evaluations. Error bars represent 95% confidence intervals.*

The second hypothesis posited that (a) participants will exhibit elevated confidence in their credibility judgement when opting to post a comment online, compared to when they refrain from posting, and (b) the act of disputing a claim will be associated with higher confidence ratings compared to instances of endorsement. As shown in figure 4, mean confidence in credibility evaluations was high in instances where participants decided to give comments online as opposed to abstaining. Further, the mean confidence rating of the initial evaluation behind refuting comments was higher than that of endorsing ones. A Welch's (ANOVA) test with Games-Howell post-hoc analysis was used to test the second hypothesis since Levene's test for the homogeneity of variance was significant for the confidence ratings given by the participants *(F (3, 2838) = 18.849, p < 0.001)*. The analysis was conducted for four groups: 2 comment types (disputing/ endorsing) x 2 online response conditions (comment given online/ not given online). The Welch's test showed a significant difference in confidence *(F (3.00, 951.97) = 100.52, p < .001)* between the four groups. Games-Howell post-hoc analysis revealed that the mean confidence for disputing comments given online *(N= 323, M= 78.06, SD=25.68)* was significantly higher than that of disputing comments *(N= 1187, M= 58.65, SD=29.48, p < 0.001)* and endorsing comments *(N= 990, M= 53.06, SD=25.34, p < 0.001)* not given online. Similarly, participants showed significantly higher confidence in their credibility evaluations when they gave endorsing comments online *(N= 342, M= 70.54, SD=23.76)* than disputing comments *(N= 1187, M= 58.65, SD=29.48, p < 0.001)* and endorsing comments *(N= 990, M= 53.06, SD=25.34, p < 0.001)* not given online. Furthermore, participants expressed significantly higher confidence when giving disputing comments *(N= 323, M= 78.06, SD=25.68)* compared to endorsing comments online *(N= 342, M= 70.54, SD=23.76, p< 0.001)*. Therefore, both H2a and H2b were supported.

Hypothesis three posited that participants would exhibit greater discernment when providing disputing online comments compared to endorsing ones, such that the "perceived false" evaluations behind disputing comments would be more accurate (more frequently match the ground truth) than the "perceived true" evaluation behind endorsing comments. In other words, if comments where the credibility evaluation matches the ground truth were considered as 'hits' and those that didn't as 'misses', we expected to see a higher hit ratio for disputing comments compared to endorsing comments. Figure 5 depicts the proportion of evaluation accuracies between claims perceived as true or false and those that received comments online or not within those perception groups. A two-sample test for equality of proportions with continuity correction revealed that the proportion of hits for disputing comments (78.95%) was significantly greater than that of endorsing comments (68.13%), providing support for H3 *($\chi^2$ (1) = 9.408, p = .002)*.

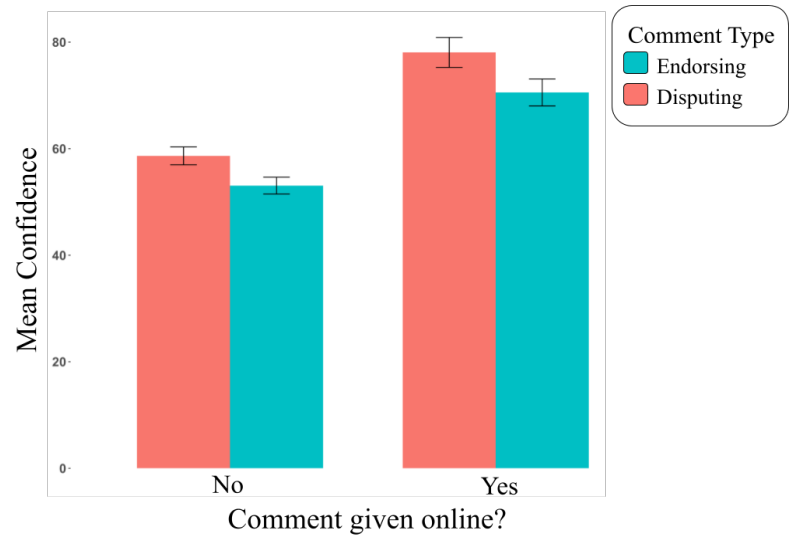

*Figure 4.   Mean confidence ratings between four groups: 2 comment types (disputing/ endorsing) x 2 online responses (given online/ not given online). Error bars represent 95% confidence intervals.*

Further, when considering the accuracy of the credibility evaluations of the claims that did not receive comments online, we found that participants had significantly fewer hits in the perceived false group (64.53%) compared to the perceived true group (71.21%, $\chi^2$ (1) = 10.688, p = .001). Notably, there was no significant difference in hits *($\chi^2$ (1) = 1.0169, p = 0.31)* between perceived true claims that were given endorsing comments online (71.21%) and those that did not get endorsing comments (68.12%). In contrast, however, credibility evaluations behind disputing comments given online (64.53%) had a significantly higher proportion of hits *($\chi^2$ (1) = 1.0169, p < 0.001)* compared to those that did not get disputing comments online (78.94%) (see Figure 3 graph B). A potential reason behind the inaccuracy of perceived false claims that were not commented on might be that participants were more inclined to perceive claims as false more often than true, as revealed by a chi-square test *($\chi^2$ (1) = 11.148, p < 0.001)*.

In an additional analysis, following Batailler et al. (2022)'s approach of using SDT in the identification of misinformation on social media, we compared the discernment ability and response bias of credibility evaluations of claims that received comments online and those that did not. A Student's test revealed that there was no significant difference *(t(243) = -0.59, p = .554, $CI_{95}$=[-0.23, 0.12])* between the discernment ability of evaluations behind claims that were commented on *(m=.89)* and not *(m=.83)*. However, participants showed a significant response bias *(t(243) = 3.62, p < 0.001, $CI_{95}$=[ 0.08, 0.27])* such that they were more likely to perceive claims to be misinformation among the claims that were not commented on *(m=.11)* compared to those that were *(m=-0.06)*. This would mean that participants were no more likely to accurately differentiate between true and false claims that were commented on compared those that were based on SDT's indices. This suggests that they did not spend more cognitive effort when evaluating the credibility of the claims they commented on. Additionally, within the group that didn't receive comments, participants showed a higher tendency to perceive claims to be misinformation while those that did get comments had a higher tendency to be perceived true.

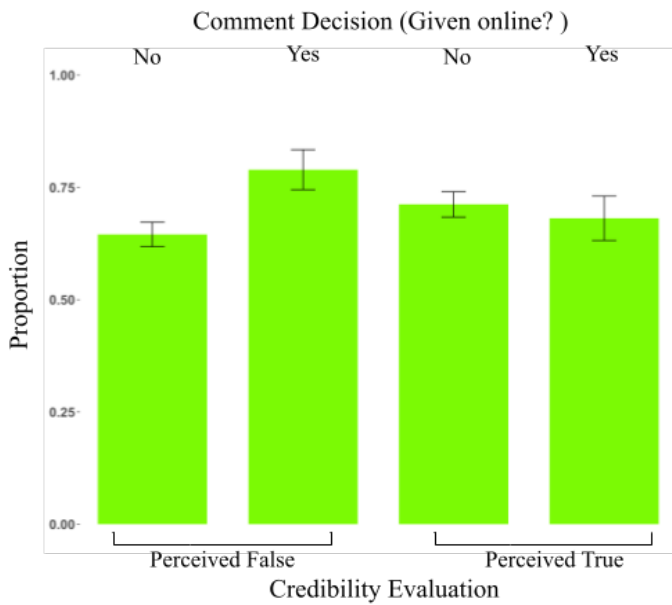

*Figure 5.    Distribution of the proportions of 'hits' between the two groups of credibility evaluations and comment decisions (comment given online/ not given online). Error bars represent 95% confidence intervals.*

# 5    Discussion

Social media users often rely on comments given by other users when assessing the credibility of a post (Geeng et al., 2020). Prior research has found that comments can influence how users perceive the credibility of the posts they accompany (Ransom et al., 2021; Wijenayake et al., 2021). Importantly, social correction – corrections given by other users to misinformation seems to have the potential to mitigate misperceptions among social media users (Bode & Vraga, 2018; Sullivan, 2019; Vraga & Bode, 2017, 2018). Despite the growing attention in the literature towards motivating social media users to give corrections online, little is known about the relative frequency and reliability of disputing or endorsing comments. The current study investigated how reputational concerns may have an asymmetric effect on whether people are prepared to post endorsing or disputing comments given their credibility evaluation of a post and their confidence in that evaluation.

When perceiving credibility online, social media users have been found to have a tendency to evaluate posts as misinformation more often than not in experimental settings, indicating a deception bias in their evaluations (M. Luo et al., 2022). Our findings were consistent with this, as participants showed a higher tendency to perceive claims as misinformation than true posts. Despite exhibiting such a deception bias in our study, participants gave relatively fewer disputing comments compared to endorsing ones. This discrepancy can be attributed to participants' perception that the impression management outcomes achieved through disputing others' claims are less valuable or desirable than those attained via endorsement. Previous research suggests that users are concerned about disputing comments offending other users and sparking online arguments (Gurgun et al., 2024; Hermansyah et al., 2021; T. Johnson & Kromka, 2023). Therefore, participants likely took a conservative approach when correcting as they may be less motivated to manage their online impressions through disputing comments.

When individuals assess how they construct their desired impression, they consider the probability of successfully achieving the intended impression and the potential reputational costs if they fail to do so (Leary & Kowalski, 1990). Given that users want to make a positive impression of themselves with others through their online actions, it is likely that they would consider if the comment they give can potentially harm their reputation in any way. Hence, users would want to avoid undesired identities, such as being perceived as someone who lacks judgment by giving comments that misalign with the ground truth. In line with this, our study found that the cases where participants decided to give comments online coincided with higher confidence levels in their initial credibility evaluation. Further, we observed that choosing to post a disputing comment rather than an endorsing comment also coincided with higher confidence. We take this as evidence that people need to be more confident in their judgment when deciding to post as opposed to merely making a judgment decision on a post's credibility, and even more so when posting a disputing comment. This requirement for heightened confidence is likely attributable to the distinct reputational risks associated with each type of comment. While a mismatched endorsing comment generally poses reputational risks, a mismatched disputing comment that directly and publicly challenges the credibility of another person's post can potentially incur greater reputational costs.

Given the asymmetric reputational costs of a mismatched disputing comment compared to a mismatched endorsing comment, we found evidence that suggests social media users would be more conservative when giving disputing comments as opposed to endorsing. According to Lynn & Barrett (2014), individuals adopt a more conservative criterion when the cost of a miss is perceived to be high and as a result, they want to minimize the probability of misses occurring (Lynn & Barrett, 2014). In line with this, we expected to see a higher accuracy rate (hit rate) among disputing comments than endorsing ones. In fact, we found that participants were more discerning when disputing others, such that they relatively gave disputing comments to actual misinformative posts more often than they gave endorsing comments to actual true posts. Therefore, the findings from our study suggest that although social media users tend to take a more conservative approach, which results in relatively fewer disputing comments, those comments are generally more reliable than endorsing ones.

## 5.1  Theoretical contribution

Existing literature explores how social media users engage in the behavioural process of managing their online impressions when sharing their emotions (Bazarova et al., 2015), travel selfies (Lyu, 2016), mass media content (B. K. Johnson & Ranzini, 2018), and when trying to impress potential employers (Roulin & Levashina, 2016), all in the hopes of gaining a positive online reputation from others. While previous research has examined the motivations and constraints influencing the provision of corrective comments on social media (Bautista et al., 2021b; Gurgun et al., 2024; Koo et al., 2021; C. Luo et al., 2024), little is known about the frequency and reliability of such comments. To our knowledge, this study represents the first comprehensive investigation of these factors.

Utilising the two-component model of impression management (Leary & Kowalski, 1990), we examine the frequency with which social media users provide disputing and endorsing comments, taking into account their online reputational concerns. Furthermore, we investigate how the differential criteria adopted by users, due to the asymmetric perceived value of each comment type, lead to varying levels of accuracy based on signal detection theory (Green & Swets, 1966). This approach allows us to explore the interplay between users' impression management strategies and the accuracy of their online interactions.

We find evidence that suggests users' credibility evaluation and their confidence affect the two processes of impression motivation and impression construction, which leads to different frequencies and levels of accuracy between the two types of comments. This research provides important theoretical contributions as we provide empirical evidence for the interplay between impression motivation, impression construction, and the accuracy of online interactions in the context of social media credibility evaluations.

## 5.2  Practical Implications

Our study provides further evidence that social correction can be a viable intervention against misinformation on social media platforms. Though less frequently given than endorsing comments, disputing comments proved to be more reliable. Social media service providers should identify ways to motivate users to give corrections online as most users believe that these platforms do not provide the necessary affordances to challenge misinformation (Gurgun et al., 2023). Since user activity on such platforms is primarily driven through positive incentives such as likes and upvotes (Bazarova et al., 2015), new features incentivizing taking corrective action against misinformation may yield positive results such as user rating systems that reward users for accurate and helpful corrections, incentivizing positive behavior. Additionally, considering the recent advancements in generative ai, social media providers can potentially utilize large language models combined with retrieval-augmented generation to assist users in crafting polite and evidence-based corrections. This may also provide an opportunity to overcome the limited scalability issue of professional factchecking organizations as users actively engage in correcting misinformation. Moreover, there may be opportunities to improve the detection accuracy of deep learning models designed to detect misinformation based on the disputing comments given by users (Kim & Walker, 2020).

## 5.3     Limitations and future directions

The current study is not without limitations. First, the sample of participants in this study were undergraduate students. As prior studies have shown, younger social media users may be more willing to give disputing comments to misinformation than older counterparts (Heiss et al., 2023; Huber et al., 2022). Therefore, future studies should investigate whether these findings generalise to more diverse populations.

While this study examined how individuals dispute and endorse social media content within a controlled, short-term experimental framework, future research should leverage longitudinal social media data analysis to better capture the dynamic nature of user interactions; furthermore, though we observed that higher confidence in credibility evaluations correlates with increased online commenting, particularly disputing, the underlying factors driving this confidence require further examination.

Another important factor related to corrections is the comment content. Corrective comments are believed to be more effective if the comment's tone is not insulting (Bode et al., 2020), contains sources and is factual (Vraga & Bode, 2017). Therefore, future studies should further investigate the textual content of the disputing and endorsing comments and whether such user-generated comments can achieve their intended purpose.

From an impression management perspective, different individuals and even different cultures would have different levels of need for other's approval (Leary & Kowalski, 1990). Therefore, future studies should explore how the need for others' approval influences users' commenting behaviour. Our study found that despite users' best efforts, there can be instances where they give disputing comments to true information and endorse misinformation. Such cases can harm their online reputation, and they might engage in reputation-repairing behaviour afterwards. Nevertheless, how they would do so is yet to be explored, as users may give excuses for their mistakes, denigrate other users, apologise or distance themselves from the situation.

Given our findings indicating that disputing comments may exhibit higher accuracy, a logical extension of this research would be to investigate the perceptions of the users receiving them. Specifically, future studies should examine whether social media users recognise the potentially higher reliability of disputing comments compared to endorsing ones. Ideally, users should be calibrated to accurately assess the relative reliability of different comment types. Such calibration could significantly enhance the efficacy of user-generated corrections and contribute to more informed decision-making in online environments.

## 6     Conclusion

Using the two-component model of impression management and the signal detection theory as the theoretical lens, this study explored how social media users' credibility evaluations of content they encounter are reflected in their commenting behaviour. The results suggest that users take a more conservative approach when giving disputing comments as opposed to endorsing comments. Nevertheless, users displayed better discernment when disputing compared to when endorsing another user's post. These findings provide further evidence for the fact that social correction can be a viable intervention against mitigating misinformation on social media platforms.